\documentclass[
 reprint,
 superscriptaddress,
 amsmath,amssymb,
 aps,
 prb,
]{revtex4-2}

\usepackage{graphicx}
\usepackage{dcolumn}
\usepackage{bm}
\usepackage{braket}
\usepackage{physics}
\usepackage{ulem}
\newcommand{\ve}[1]{{\mbox{\boldmath $#1$}}}

\usepackage[unicode=true]{hyperref}
\usepackage{xcolor}
\definecolor{ori_blue}{HTML}{2E3092}

\hypersetup{
  colorlinks = true, 
  linkcolor = blue, 
  filecolor = blue, 
  urlcolor = blue, 
  citecolor = blue 
}
\begin{document}


\title{Anomalous enhancement of Néel order in the $S = \frac{1}{2}$ square lattice Heisenberg model under fictitious magnetic field
}

\author{Takayuki Yokoyama}
\affiliation{Quantum Matter Program, Graduate School of Advanced Science and Engineering, Hiroshima University,
Higashihiroshima, Hiroshima 739-8530, Japan}

\author{Yasuhiro Tada}
\email[]{ytada@hiroshima-u.ac.jp}
\affiliation{Quantum Matter Program, Graduate School of Advanced Science and Engineering, Hiroshima University,
Higashihiroshima, Hiroshima 739-8530, Japan}
\affiliation{Institute for Solid State Physics, University of Tokyo, Kashiwa 277-8581, Japan}

\begin{abstract}
Fictitious magnetic fields can be introduced in quantum magnets by strain engineering and the Aharonov-Casher effect. 
Here, we study impacts of a uniform fictitious magnetic field and corresponding Landau quantization on the Néel order in the $S= \frac{1}{2}$ Heisenberg model on the square lattice as a prototypical system of quantum magnets.
We first analyze the system by using the spin-wave approximation. 
It is found that the staggered magnetization is enhanced by the fictitious magnetic field and it shows a simple scaling behavior with respect to the magnetic length,
in contrast with a naive expectation based on the Dzyaloshinskii-Moriya interaction. 
We also perform density matrix renormalization group calculations to fully consider quantum effects and obtain quantitatively more accurate results.
The enhancement is found to be anomalously strong and 
it shows a non trivial scaling behavior or a fluctuation induced discontinuity. 
These results are beyond the well-known understanding of correlated systems under magnetic fields called the magnetic catalysis.
\end{abstract}

\maketitle


\section{\label{sec:introduction}Introduction}
A magnetic field has significant impacts on physical properties not only through the Zeeman effect but also through the orbital effect.
The orbital effect is related to the Landau quantization of a band structure and it
has been studied mainly for itinerant electron systems.
However, a fictitious magnetic field and corresponding Landau quantization can be introduced in electrically charge-neutral insulators.
For example, a fictitious magnetic field can be introduced by strain.
The strain-induced fictitious magnetic field was discussed for graphene, where triaxial strain can effectively lead to a non-uniform magnetic field~\cite{Guinea2010,Levy2010}.
Similar mechanisms also work for electrically charge-neutral quasi-particles such as emergent Majorana fermions and bosonic magnons in quantum spin systems on the honeycomb lattice, because strain generally leads to modulation of hopping integrals, and electromagnetic coupling is not necessary~\cite{PhysRevLett.116.167201,PhysRevLett.123.207204,PhysRevB.99.214413,PhysRevResearch.3.043223, PhysRevB.104.125117}.
The strain-induced fictitious magnetic fields were generalized to other lattices such as kagome, $\alpha$-$T_3$, and square lattices~\cite{PhysRevB.102.045151,PhysRevB.106.155417,PhysRevB.106.245106,PhysRevB.108.205149}. 

The other promising approach for realization of a fictitious magnetic field in a quantum magnet is the Aharonov-Casher effect~\cite{PhysRevLett.53.319,avishai2023}.
In a system with a spin-orbit interaction, an electric field adds to momentum of electrons and thus can behave as an SU(2)  gauge field in the spin space, which modifies magnetic interactions and leads to Dzyaloshinskii-Moriya like interactions.
Such an effect has been evaluated theoretically~\cite{PhysRevLett.106.247203,PhysRevB.105.174411} and examined experimentally for magnets~\cite{PhysRevLett.113.037202,PhysRevB.108.L220404}.
The fundamental difference between the strain engineering and the Aharonov-Casher effect is that a spatially uniform fictitious magnetic field can be realized and spin-rotation symmetry is reduced in the latter.
For a one-dimensional spin chain, the Aharonov-Casher effect with a uniform electric field is simply implemented by the twisted boundary condition, and field dependence of the spin current in the Heisenberg model was studied using the Bethe ansatz~\cite{PhysRevB.52.6569}.
In two dimensions, a uniform electric field generates a constant Aharonov-Casher phase in the magnetic interaction, while an electric field gradient can induce a fictitious magnetic flux which leads to Landau quantization of magnons~\cite{PhysRevB.95.125429,PhysRevB.96.224414,OWERRE201893}.
Theoretically, within the spin-wave approximation neglecting magnon interactions, it was shown that magnons under a fictitious magnetic field can have topologically non-trivial bands and exhibit various Hall effects.
It was also proposed that a fictitious magnetic field can give rise to the spin Nernst effect of magnons based on the spin-wave approximation~\cite{PhysRevLett.125.257201}.
The basic physics behind these interesting transport phenomena is controlling magnon bands, which is called magnonics~\cite{MagnonReview}.
In this context, the fictitious magnetic field induced by the Aharonov-Casher effect can be a useful tool for engineering magnons.

However,
despite extensive studies of quantum spin systems with fictitious magnetic fields, their impacts on magnetic orders have not been well understood.
It is a non trivial problem whether the magnon interactions and quantum fluctuations often ignored within the spin-wave approximation are indeed safely negligible.
It was shown that, in the $S = \frac{1}{2}$ Heisenberg model on the honeycomb lattice under a strain-induced non uniform field, the antiferromagnetic order is weakened near the corners of the system and suppression in the quantum Monte Carlo calculations is much stronger than that in the spin-wave approximation, implying the importance of quantum fluctuations~\cite{PhysRevB.104.125117}.
On the other hand, to our knowledge, magnetism under a uniform fictitious magnetic field has not been theoretically explored, and the fate of the magnetism is not known.
In contrast with strain-induced fields, a fictitious magnetic field by the Aharonov-Casher effect reduces the spin rotation symmetry and thus a magnetic order could show different behaviors from those under strain.
For an electric field which is uniform within a plane, the corresponding fictitious vector potential can be approximated by a Dzyaloshinskii-Moriya interaction, and it will suppress the Néel order.
In the presence of an electric field gradient which gives a fictitious magnetic flux, however, validity of such an approximation is subtle and enhancement of the Néel order might be possible.
Furthermore, a change of the magnetic order could be strengthened by magnon interactions.

In this paper, we discuss effects of uniform fictitious magnetic fields in the $S= \frac{1}{2} $ Heisenberg model on the square lattice as a prototypical model of quantum spin systems. 
We first analyze the model using the spin-wave approximation, and next perform density-matrix-renormalization-group (DMRG) calculations to fully take consider quantum effects into account and obtain quantitatively more accurate results~\cite{PhysRevLett.69.2863, annurev-conmatphys-020911-125018, 10.21468/SciPostPhysCodeb.4}.
It is found that the Néel order is enhanced by the fictitious magnetic field in contrast with the above-mentioned naive expectation based on the uniform Dzyaloshinskii-Moriya interaction.
Furthermore, the DMRG calculation shows much stronger enhancement of the Néel order, 
which indicates the importance of the quantum fluctuations neglected in the spin-wave approximation.
We point out that this enhancement could be regarded as a bosonic variant of the magnetic catalysis, which has been extensively studied , mainly for Dirac fermion systems~\cite{MIRANSKY20151,PhysRevLett.73.3499,PhysRevD.52.4718,PhysRevD.46.2737,PhysRevB.85.195417,PhysRevD.107.094505,GUSYNIN1999361,PhysRevB.66.045108,JHEP07(2011)037,PhysRevResearch.2.033363, Semenoff2011}.
However, the anomalous enhancement has not been seen in any fermion systems and is beyond the standard magnetic catalysis.
The fictitious magnetic field in quantum magnets could provide an interesting platform for the interplay of Landau quantization and interactions.

\section{\label{sec:model}Model and Calculation Method}
\begin{figure}[tb]
     \centering
     \includegraphics[width=0.85\columnwidth]{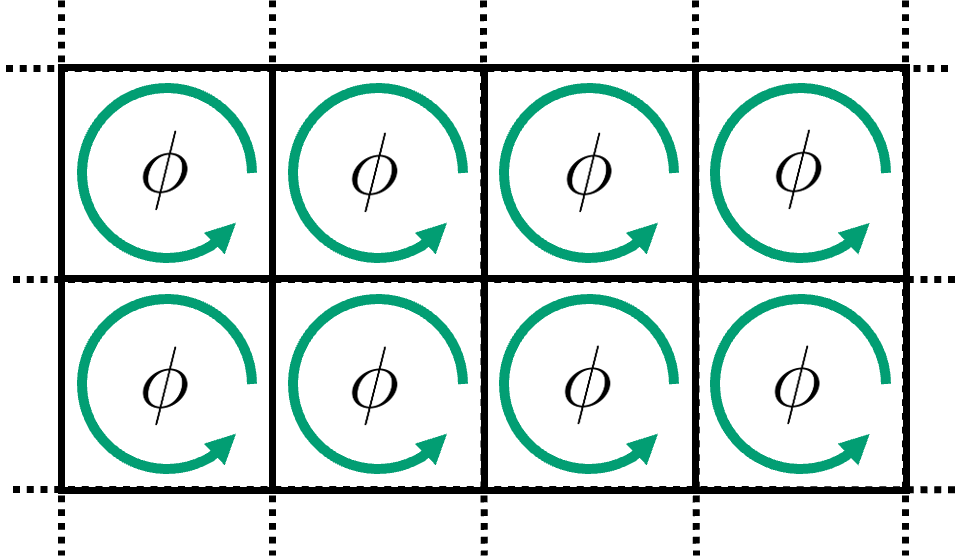}
     \caption{A schematic picture of the square lattice under the fictitious magnetic flux $\phi$ per plaquette. }
     \label{flux}
\end{figure}

We study the $S = \frac{1}{2}$ antiferromagnetic Heisenberg model on a square lattice (system size $L_x\times L_y$) under a fictitious magnetic flux $\phi$, as shown in Fig.~\ref{flux}.
Although the fictitious magnetic field in the Aharonov-Casher effect is generally described by an SU(2) gauge field, we focus on the U(1) part in our analysis. 
The magnetic field is uniform over the entire system, $\sum_{ij \in p} \theta_{ij} = 2 \pi \phi$ for every plaquette $p$. 
Similar to the Aharonov-Bohm effect, the spin  acquires a phase  $2 \pi \phi$ when it goes around a plaquette.
Note that the lattice constant is the length unit in our model, and there is no distinction between a magnetic field and a magnetic flux per plaquette, nor between the vector potential and the phase factor.
The fictitious magnetic field could be realized by an electric field gradient within the plane through the Aharonov-Casher effect or alternatively by an inhomogeneous Dyzaloshinskii-Moriya interaction~\cite{PhysRevB.95.125429,PhysRevB.96.224414,OWERRE201893,PhysRevLett.125.257201,MagnonReview}.
The Hamiltonian is 
\begin{equation}
\mathcal{H} = J
\sum_{\ev{i,j}} \left[ \frac{1}{2} \left( e^{i {\theta}_{ij}} S^{+}_i S^{-}_j + e^{-i {\theta}_{ij}} S^{-}_i S^{+}_j \right) + S^z_i S^z_j \right],
\label{eq:spin}
\end{equation}
where $J>0$ is an antiferromagnetic interaction, and $\ev{i,j}$ is a pair of nearest-neighbor sites.
The $S_j^{\pm}$ operators are $S_j^{\pm}=S_j^x\pm iS_j^y$.
The present magnetic field is a fictitious field and the Hamiltonian does not contain a Zeeman term.
We introduce the fictitious vector potential $\theta_{ij}$ in the Landau gauge as
\begin{equation}
{\theta}_{ij} = 2 \pi \phi \frac{x_i + x_j}{2} \left(y_i - y_j \right),
\label{eq:theta}
\end{equation}
where the site position is denoted by $\bm{r}_j=(x_j,y_j)$
with $x_j=0,1,\dots,L_x-1$ and $y_j=0,1,\dots,L_y-1$. 
The Heisenberg model with the vector potential has  U(1) global spin-rotation symmetry along the spin-$z$axis, and the spin-$\pi$ rotation symmetry along the-$x$ axis combined with the complex conjugation, $z\to z^*, S_j^+\to S_j^-, S_j^z\to -S_j^z$, where $z$ is a complex constant.
Note that, in the standard notation of the spin operators with the Pauli matrices, the $y$-component transforms as $S_j^y \to -S_j^y$ under the complex conjugation.
Thus, SU(2) spin-rotation symmetry at $\phi=0$ is reduced to ${\mathrm U}(1)\times {\mathbb Z}_2$ symmetry in the presence of $\phi\neq0$ (the $\mathbb{Z}_2$ symmetry is anti-unitary).
The combined-$\pi$ rotation symmetry is spontaneously broken in the $z$-axis Néel ordered state. 
In this paper, we use two system geometries, namely a torus and a cylinder,
where there are no boundary effects in the former while the latter is convenient for discussions on small fluxes.
We impose periodic boundary conditions for both $x,y$-directions in the torus geometry, for which the flux is quantized as $\phi=p/L_x$,  with $p\in{\mathbb Z}$. 
On the other hand, the cylinder geometry has open (periodic) boundary conditions for the $x(y)$ direction.  
The flux is quantized as $\phi=p/\{(L_x-1)L_y\}$, for which (projective) mirror symmetry $x\to L_x-x$ is kept in the cylinder.

Since the Aharonov-Casher effect is a relativistic phenomenon, an experimentally achievable $\phi$ would be small.
For example, when the lattice constant is  1 $\mathrm{\AA}$ and the electric field gradient $\mathcal{E} = 200  {\rm V/ \mu m^2}$, the fictitious magnetic field is $\phi \sim 0.01$ and the magnetic length $l_B \sim 35.6 {\rm V/ \mu m}$.
On the other hand, our DMRG analysis in Sec. \ref{sec:DMRG} reveals nontrivial behavior of the Néel order parameter even under a very small fictitious magnetic field. 
This implies that it would be possible to experimentally observe
changes of the Néel order under the fictitious magnetic field.
We note that an efficient method for achieving a moderate Aharonov-Casher effect has been proposed in a previous study~\cite{PhysRevB.95.125429}.
Moreover, a fictitious magnetic field with a scale larger than the Aharonov-Casher effect could be applied to spin systems through artificial gauge fields in cold atomic systems~\cite{PhysRevLett.111.185301, PhysRevLett.111.185302}.

In this paper, we first apply the spin-wave approximation, which is a standard method for describing the low-energy properties of quantum magnets.
The Hamiltonian is approximated within quadratic terms in the boson operators and interactions between the magnons are neglected (see Appendix \ref{appendix::spin_wave}).
In the spin-wave approximation, both torus and cylinder geometries are used depending on physical quantities considered and ranges of the fictitious magnetic flux.
Next, we perform the DMRG calculation in the cylinder geometry to obtain more accurate results.
The quantum fluctuations arising from the magnon interaction neglected in the spin-wave approximation are considered in the DMRG calculation.
The aspect ratio is chosen as $(L_x,L_y)=(2L,L)$, with the system size $L=6,8, \text{and} \, 10$.
The bond dimension is $\chi = 800 -3200$, and the truncation error is $\epsilon < 10^{-5}$. 
\cite{annurev-conmatphys-020911-125018, PhysRevLett.69.2863, 10.21468/SciPostPhysCodeb.4}.

\section{\label{sec:level1}Calculation Results}

In this section, we show the calculation results of the spin-wave approximation and the DMRG at zero temperature.
They give qualitatively consistent results, but there is significant quantitative difference.
To have insight into the results of DMRG calculation , we give a brief discussion based on the magnetic catalysis.

\subsection{\label{sec:level2} Spin-wave approximation}

In the spin-wave approximation, we suppose that the classical ground state at $\phi=0$ is the Néel ordered state in the spin-$z$axis.
Under a fictitious magnetic field, the system has U(1)${\times \mathbb Z}_2$ symmetry, 
where U(1) symmetry breaking corresponds to Néel order in the $xy$-plane and ${\mathbb Z}_2$  symmetry breaking to Néel order along the $z$-axis. 
Therefore, the Néel order parameter along the $z$-axis is enhanced if 
${\mathbb Z}_2$  symmetry breaking is more strongly favored in rhe presence of a fictitious magnetic field. 
On the other hand, the staggered moment will be aligned in the $xy$-plane if U(1) symmetry is spontaneously broken by the fictitious magnetic field, 
which means that the $z$-axis Néel order will be suppressed.
In the latter case, the $z$-axis Néel ansatz is inappropriate and the spin-wave approximation based on this ansatz state would not work. More precisely, the energy spectrum of magnons will have imaginary components when the spin-wave approximation becomes invalid.
A failure of the spin-wave approximation implies instability of the Néel ordered ansatz, and a different order may be realized in the ground state.

\begin{figure*}[tb]
     \centering
     \includegraphics[width=1.0\textwidth]{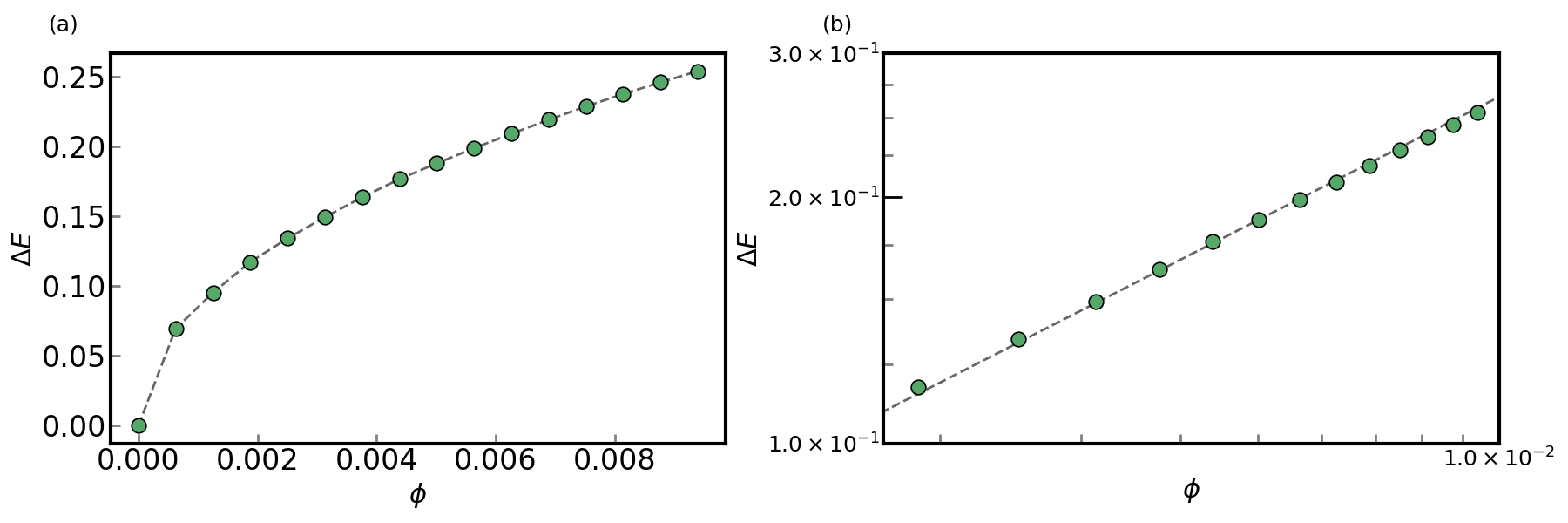}
     \caption{Magnon energy gap for small values of the fictitious magnetic field (a) in the linear scale and (b) in the logarithmic scale. The system size is $L_x = L_y = 100$ sites with the cylinder geometry. The dashed curve in (b) is $\Delta E\sim \sqrt{\phi}$. Small deviation from the exact $\Delta E\sim \sqrt{\phi}$ behavior is considered a finite-size effect.}
     \label{energy_cri}
\end{figure*}

We first calculate the energy spectrum of the magnons and investigate how the magnon excitations depend on the fictitious magnetic flux $\phi$.
\begin{figure}[b]
     \centering
     \includegraphics[width=1.0\columnwidth]{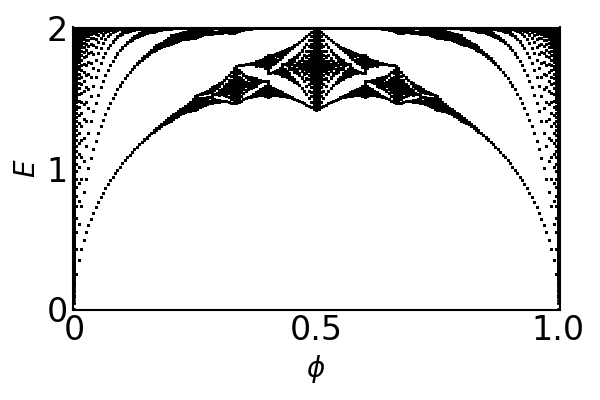}
     \caption{Magnon energy spectrum under the fictitious magnetic flux $\phi$ within the spin-wave approximation. The system size is $L_x=L_y=200$ in the torus geometry. }
     \label{hofsta}
\end{figure}
Figure~\ref{hofsta} shows the energy spectrum calculated by the spin-wave approximation for various magnetic fluxes $0\leq \phi \leq 1$.
The magnon spectrum shows a Hofstadter butterfly as in a free fermion system, as demonstrated in previous studies~\cite{PhysRevB.14.2239,OWERRE201893,PhysRevLett.125.257201}.
Note that the calculated magnon energy is always real (nonimaginary) for all $\phi$, which implies that the $z$-axis Néel ground state is stable under the fictitious magnetic field.
If the $z$-axis Néel state were unstable under some fluxes, the magnon energy would take complex values.  
One can see that the magnon excitation acquires an energy gap under the magnetic field $\phi\neq0$.
The magnon excitation gap $\Delta E$ (the smallest magnon excitation energy) behaves as $\Delta E\propto \sqrt{\phi}$ for small magnetic fluxes $0<\phi\lesssim 0.01$, as shown in Figs.~\ref{energy_cri}(a) and (b), which is consistent with the long-wavelength behavior described by the Klein-Gordon equation~\cite{PhysRevLett.125.257201}.
\begin{figure*}[tbh]
     \centering
     \includegraphics[width=1.0\textwidth]{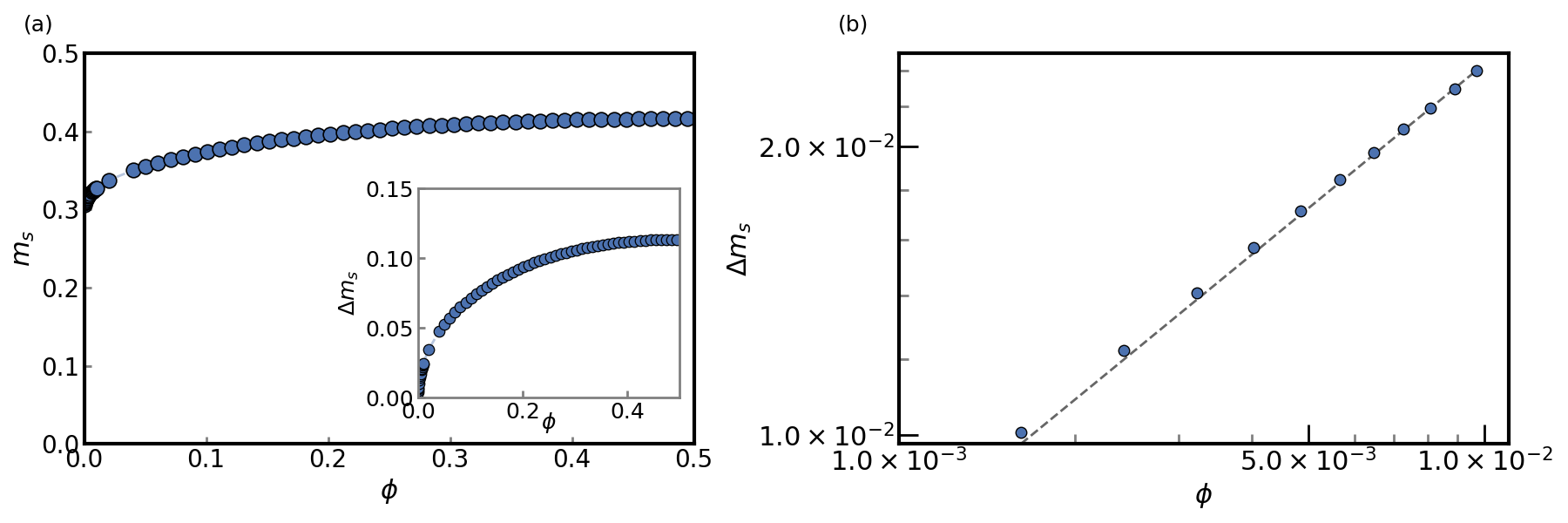}
     \caption{(a) Néel order parameter $m_{s}$ calculated by the spin-wave approximation in the cylinder geometry for the system size $L_x=100, L_y=50$. The inset is the difference $\Delta m_{s}$. (b) Logarithmic scale plot of $\Delta m_s$. The dashed line represents $\Delta m_s \propto \sqrt{\phi}$. }
     \label{neel}
\end{figure*}
The $\Delta E\propto \sqrt{\phi}$ dependence can be understood by a simple dimensional analysis of the linear spectra in the momentum space $\Delta E\propto k$ at $\phi=0$. 
Indeed, $\Delta E\propto k\propto 1/L$ for the system size $L$,
and one can just replace $L$ with the magnetic length $l_B=1/\sqrt{\phi}$ to obtain $\Delta E\propto 1/l_B=\sqrt{\phi}$.
Note that the long-wave length limit of the magnon spectrum is effectively described by the Klein-Gordon equation and there is no zero-mode ($\Delta E=0$) in sharp contrast with Dirac fermions at charge neutrality.
The non-zero magnon gap implies that any excitations from the ground state are gapped since the magnons are elementary excitations in the Heisenberg model.
The correlation length will be given by $\xi\sim v/\Delta E\propto l_B$, where $v$ is the speed of the magnon excitation, and the system is governed by this length scale.

Next, we calculate the Néel order parameter $m_{s}$ given by
\begin{align}
  m_{s} = \frac{1}{N}\sum_{j} \ev{S^z_j} (-1)^{j},
\end{align}
where we use use the cylinder geometry since it allows small fluxes.
Here $N$ is the number of sites in the summation and
it is restricted to the $N=L\times L$ sites in the middle of the system for the cylinder geometry with $(L_x,L_y)=(2L,L)$ to suppress contributions from the open boundary.
Note that the spin-wave approximation works quite well for the Heisenberg model at $\phi=0$ and it gives $m_s\simeq 0.303$ in the thermodynamic limit which is very close to $m_s\simeq 0.306 - 0.309$ obtained by the extensive numerical calculations~\cite{PhysRevLett.99.127004,PhysRevB.82.024407,BISHOP2017178,doi:10.7566/JPSJ.92.023701,PhysRevResearch.5.013132,PhysRevB.110.054411}.
Figure~\ref{neel}(a) shows the Néel order parameter $m_s$ under the fictitious magnetic flux $\phi$.
We find that the Néel order parameter is enhanced from the zero-flux value $m_s(0)\simeq 0.303$, and it reaches $m_s\sim 0.41$ at large fluxes.
Once the enhanced $m_s$ has been obtained, it can be naturally understood from the gapped magnon spectrum in analogy to an Ising-like system at $\phi=0$, where magnons are gapped, and correspondingly, $m_s$ is large.  
Since zero-point motion of the magnons leads to the suppression of $m_s$ from the classical value $m_{\rm max}=0.5$ within the spin-wave approximation, a magnon gap would generally imply increase of $m_s$.
At the same time, however, the enhanced Néel order may be non-trivial,
once one notices that the Hamiltonian in Eq.\eqref{eq:spin} partly includes a non-uniform Dzyaloshinskii-Moriya interaction in the spin $xy$-plane when $e^{i\theta_{ij}}$ is simply Taylor expanded up to the lowest order in $\phi$.
A uniform $xy$-Dyzloshinskii-Moriya interaction is known to destabilize the $z$-axis Néel order, and one might naively expect that the non-uniform Dyzloshinskii-Moriya interaction has a similar influence.
The enhancement of the Néel order in our calculation suggests that it is important to fully consider the Landau quantization beyond the naive Taylor expansion.

Interestingly, the staggered moment exhibits a scaling behavior. 
To see this, we define an excess Néel magnetization measured from $m_s$ at $\phi=0$, 
\begin{align}
    \Delta m_s (\phi) = m_s(\phi) - m_s(0),
\end{align}
where $m_s(0) = 0.303$.
As shown in Fig.~\ref{neel}(b), the difference $\Delta m_s$ behaves as $\Delta m_s\propto \sqrt{\phi}=l_B^{-1}$ for small magnetic fluxes.
Generally, a physical quantity could be expanded in integer powers of $l_B^{-1}$ (with possible logarithmic corrections) within the spin-wave approximation, which is a free bosons description, since $l_B$ is the cut off scale for the gapped magnons, as mentioned above.
The $\Delta m_s\propto l_B^{-1}$ dependence is the leading-order behavior with respect to $l_B$.
In this sense, $\Delta m_s$ can be regarded as a correction to $m_s(0)$ due to the finite magnetic length $l_B$. just like the conventional finite size effect for a system with a linear size $L$~\cite{PhysRevB.39.2608}.  
In Appendix \ref{appendix::scaling_sw}, we extend our discussion to the $XXZ$ model with Ising anisotropy and demonstrate that the system is indeed governed by the magnetic length $l_B$, based on a finite-size scaling analysis.
Note that $\Delta E$ and $\Delta m_s$ slightly deviate from the exact power-law with the exponent $\frac{2}{2}$ in the very small region of $0 \leqq \phi \lesssim 0.01$.
Two possible reasons for this deviation can be considered. 
The first one is that the magnetic length $l_B$ may exceed the system size $L$, potentially leading to noncritical behaviors in the weak magnetic field region.
The second is a technical issue regarding instability in the spin-wave calculation due to specific modes which are connected to the gapless modes at $\phi=0$. 
The deviation is not physically meaningful, because these problems can be avoided in the thermodynamic limit. 

From the spin-wave calculation of the $z$-axis Néel order, one can see that the fictitious magnetic field effectively enhances the Ising interaction $J S^z_iS^z_j$.
One can compare the results above for the Heisenberg model at $\phi\neq0$ with the previous results for the $XXZ$ model at $\phi=0$, since these two models have the same symmetry ${\mathrm U}(1)\times{\mathbb Z}_2$.
The $XY$ interaction is denoted by $J_{xy}$ and the Ising interaction by $J_z\geq J_{xy}$.
In the $XXZ$ model at $\phi=0$, the spin-wave approximation gives $\Delta E(J_z)\propto \sqrt{J_z-J_{xy}}$ and $\Delta m_s(J_z) \propto \sqrt{J_z-J_{xy}}$ in the leading order~\cite{PhysRevB.44.11869,PhysRevB.46.6276}. 
The $\sqrt{J_z-J_{xy}}$ dependence in these quantities has been confirmed by various methods~\cite{PhysRevResearch.5.013132,PhysRevB.110.054411}, 
and one would naively expect that the $\sqrt{\phi}$ dependence in the Heisenberg model at $\phi\neq0$ holds beyond the spin-wave approximation as well.
However, magnon interactions are neglected in the spin-wave approximation, and validity of this approximation may be nontrivial in the present system.
To resolve this problem, we will discuss the DMRG calculations in the next section.
\begin{figure}[tb]
    \centering
    \includegraphics[width=1.0\columnwidth]{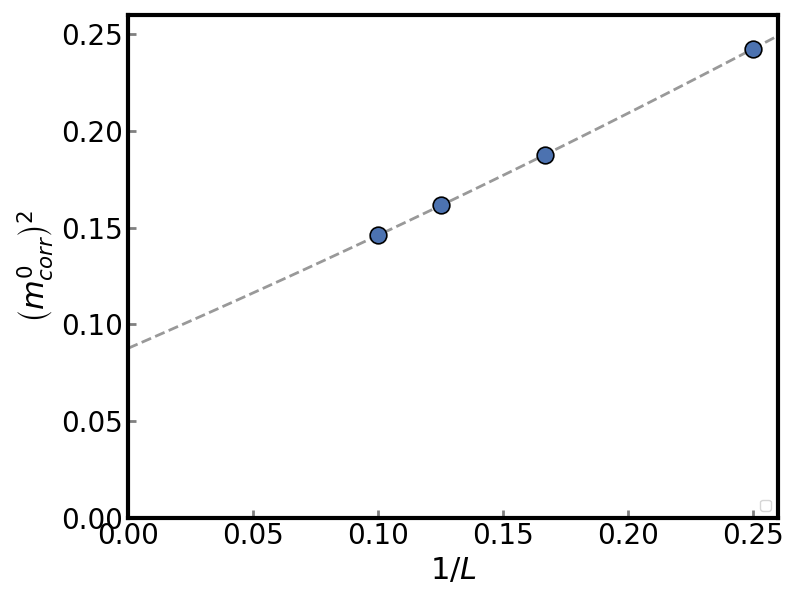}
    \caption{$(m^0_{\rm corr})^2$ plotted vs $1/L$ for $L =4, 6, 8, and \, 10$.
    The dashed curve is a polynomial fitting, $m=m_0+m_1/L+m_2/L^2+m_3/L^3$, which gives $\left(m_0 \right)^2 = 0.087(1)$ and $\left(m_0 \right) = 0.295(3)$.}
    \label{L_to_inf}
\end{figure}

\subsection{\label{sec:DMRG}DMRG calculation}
In the DMRG calculation, we focus on the correlation function for the $z$-axis Néel order,
\begin{align}
 \left(m^z_{\rm corr} \right)^2 &= \frac{1}{N^2} \sum_{j,k} \ev{S^z_j S^z_k} e^{i \ve{q} \cdot \left( \ve{r}_j - \ve{r}_k \right)},
 \label{eq:corr}
\end{align}
where $\ve{r}_j=(x_j,y_j)$ and $\bm{q} = (\pi, \pi)$.  
The summation is restricted to the $N=L\times L$ sites in the middle of the $(L_x,L_y)=(2L,L)$ cylinder system, which is a method commonly used in DMRG analysis of the two-dimensional quantum spin models to suppress contributions from the open boundaries.
We first evaluate $(m^z_{\rm corr})^2$ in the absence of the magnetic field.
The system is SU(2) symmetric at $\phi=0$, and thus, the Néel correlations are isotropic, $(m^x_{\rm corr})^2=(m^y_{\rm corr})^2=(m^z_{\rm corr})^2\equiv (m_{\rm corr}^0)^2/3$, where $(m^{x,y}_{\rm corr})^2$ are defined like in Eq.~\eqref{eq:corr}.
We show $(m_{\rm corr}^0)^2$ at $\phi=0$ in Fig.~\ref{L_to_inf} for system sizes $L=4,6,8, \text{and} \,10$.
Note that it is commonly assumed that the correlation function in the symmetry-preserving ground state coincides with the squared order parameter in the symmetry-broken ground state.
The polynomial fitting of the results for the finite size systems gives $m_{\rm corr}^0=0.295(3)$ in the thermodynamic limit $L \to \infty$.

The obtained $m_{\rm corr}^0$ is slightly smaller than the previous results $m_{\rm corr}^0=0.306 - 0.308$ evaluated by several numerical methods, ~\cite{PhysRevLett.99.127004,PhysRevB.82.024407,BISHOP2017178,doi:10.7566/JPSJ.92.023701,PhysRevResearch.5.013132,PhysRevB.110.054411}, because the system sizes used in our extrapolation are limited to $L=4 -10$. 
Nevertheless, the deviation is at most few percents
and we have confirmed that it has only negligible influence in the following discussions for $\phi\neq0$ (see Appendix \ref{appendix::QMC and DMRG}).
\begin{figure}[tb]
     \centering
     \includegraphics[width=1.0\columnwidth]{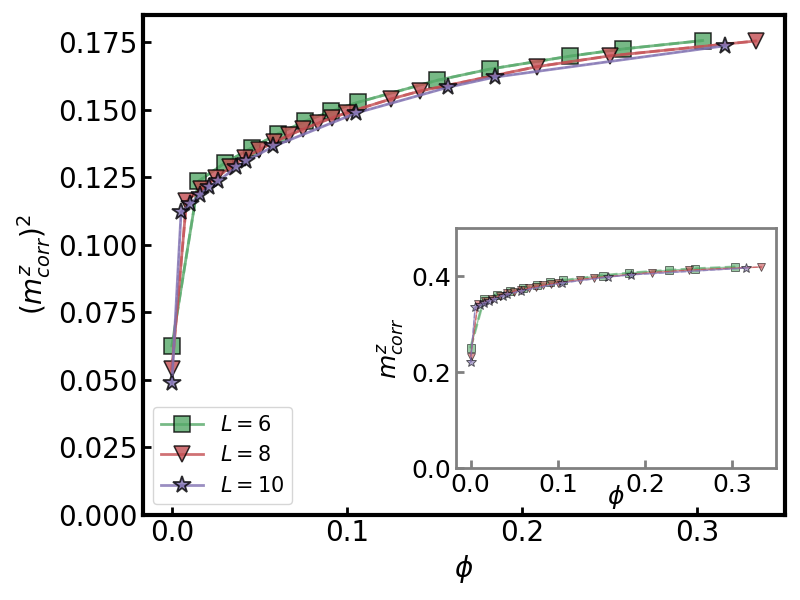}
     \caption{Néel correlation function $\left(m^z_{\rm corr} \right)^2 $ as a function of $\phi$ for vs system sizes $(L_x,L_y)=(2L,L)$ calculated by DMRG.
     The green square is for $L=6$, and the red triangle is for $L=8$,
     the purple star is for $L=10$.}
     \label{neel_dmrg}
\end{figure}
We find that the calculated ground state at $\phi\neq0$ is essentially a uniform state with the long-range Néel order like the results within the spin-wave approximation in the previous section, although there are some deviations around the open boundaries.
The ground state belongs to the $S^z_{\rm tot}=\sum_iS^z_i=0$ sector.

Figure~\ref{neel_dmrg} shows the calculated $(m^z_{\rm corr})^2$ as a function of the fictitious magnetic flux $\phi$.
Like the results by the spin-wave approximation in the previous section, the Néel magnetization is enhanced by the flux.
Additionally, the system size dependence becomes smaller as $\phi$ increases, which is qualitatively consistent with the finite correlation length $\sim l_B$ of the gapped magnons in the spin-wave approximation since finite size effects are generally suppressed in a gapped system.
However, the two results obtained by the spin-wave approximation and the DMRG are significantly different in a quantitative manner. 
Namely, the enhancement of the Néel magnetization around $\phi=0$ in the DMRG calculation is much stronger than that in the spin-wave approximation. 
This behavior could be described by a power-law function with a small exponent or by a discontinuous function.

\begin{figure}[tb]
    \centering
    \includegraphics[width=1.0\columnwidth]{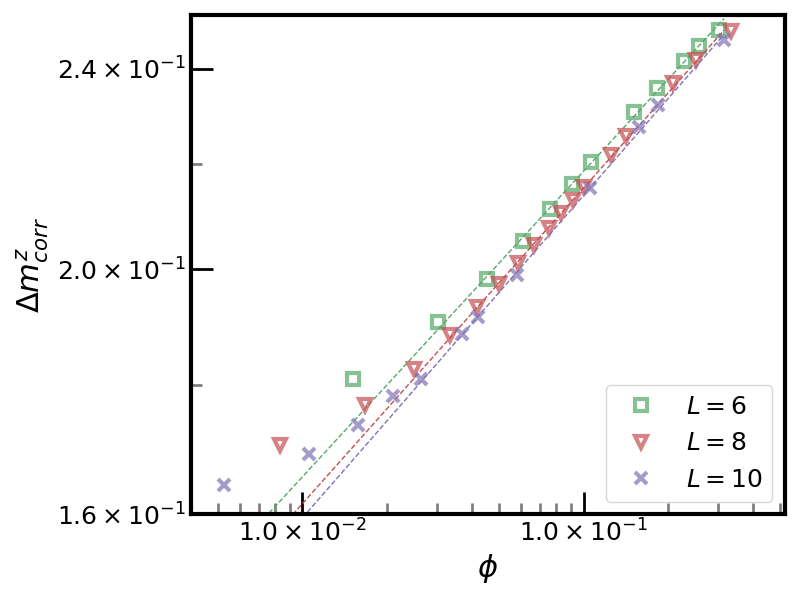}
    \caption{Scaling plot of $\Delta m_{\rm corr}^z (\phi)$ with $L = 6,8,10$.
    The power law fitting is represented by the dashed lines. }
    \label{log-log}
\end{figure}

\begin{figure}[t]
    \centering
    \includegraphics[width=1.0\columnwidth]{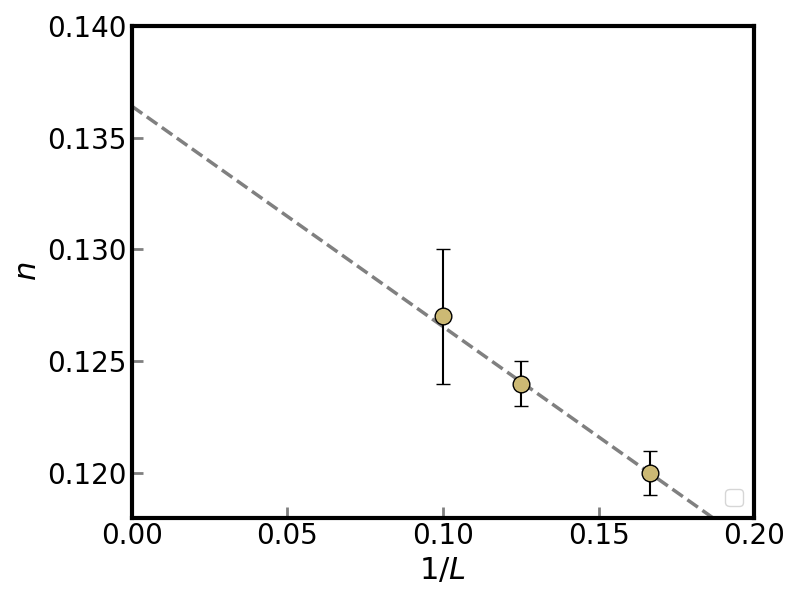}
    \caption{The exponent $n$ plotted vs $1/L$ with $L= 6,8, \text{and} 10$.
    The dashed line represents the linear fitting with weighted least squares, which gives $n\simeq 0.136$ in the thermodynamic limit.}
    \label{exponent}
\end{figure}

To understand detailed behaviors under the assumption that $m^z_{\rm corr}(\phi)$ is continuous at $\phi=0$, we introduce the difference 
\begin{align}
 \Delta m^z_{\rm corr}(\phi) = m^z_{\rm corr}(\phi) - m^z_{\rm corr}(0), 
\end{align}
where $m^z_{\rm corr}(0)=m^0_{\rm corr}(0)/\sqrt{3}\equiv 0.295/\sqrt{3}$. 
Note that $\Delta m^z_{\rm corr}(0)=0$ in the thermodynamic limit by definition
(see also Appendix~\ref{appendix::QMC and DMRG}).
We show a logarithmic scale plot of the difference $\Delta m^z_{\rm corr}$ in Fig.~\ref{log-log}.
It has only small $L$ dependence for relatively large magnetic fields as already mentioned.
In this field region,
$\Delta m^z_{\rm corr}$ exhibits a scaling behavior,
\begin{align}
    \Delta m^z_{\rm corr}=C  \phi^n,
    \label{eq:corr_fit}
\end{align}
where the prefactor $C$ only weakly depends on the system size $L$.
This scaling behavior holds above a crossover scale $\phi_L$ for each system size, and it is roughly estimated to be $\phi_6\simeq0.05, \phi_8\simeq 0.04, \text{and} \phi_{10}\simeq 0.03$.
We obtain the exponent $n$ and the prefactor $C$ for each system size $L$, and they are extrapolated for $L \to \infty$ (Fig.~\ref{exponent}). 
These results are summarized in Table~\ref{tab:powerlaw}.
Numerically, the crossover scale $\phi_L$ decreases as the system size increases
$\phi_6>\phi_8>\phi_{10}$.
Although we naively expect that $\phi_L\to 0$ as $L\to\infty$, it could converge to some nonzero value $\phi_{\infty}$, and a different scaling exponent might be obtained for $\phi<\phi_{\infty}$.
In any case, the obtained exponent $n\simeq 0.136$ for $L=6,8,10$ gives an upper bound since the slope in Fig.~\ref{log-log} for $\phi<\phi_0$ is smaller than that for $\phi>\phi_0$.
Therefore, we conclude that the fictitious magnetic field leads to the anomalously strong enhancement of the Néel order with the scaling behavior $\Delta m^z_{\rm corr}\propto \phi^{n}$ with $n\lesssim 0.136$.
Note that the above discussion is based on the assumption that the enhancement of $m^z_{\rm corr}(\phi)$ is continuous and can be described by a power-law function. 
The successful power-law fitting implies that this is indeed the case, and the $\phi$ dependence in our system is distinguished from the $J_z$ dependence of the $XXZ$ model.
If $m^z_{\rm corr}(\phi)$ is discontinuous at $\phi=0$, the enhancement is even stronger than any power-law behavior.
Such  possible discontinuous behavior may be understood as a fluctuation-induced discontinuity.
In any case, our DMRG calculation suggests the importance of the magnon interactions neglected in the spin-wave approximation.
This may be somewhat counterintuitive, because magnon interactions are usually negligible at low temperatures where there are few thermally excited magnons. 
Indeed, the significant difference between the spin-wave results and the DMRG results is highly nontrivial even when compared with the magnetic catalysis in any known systems, as discussed in the next section.
\begin{table}[t]
 \centering
 \begin{center}
  \begin{tabular}{|c|c|c|}
     \hline
    System size $L$& Exponent $n$& Prefactor $C$\\ \hline\hline
    $L=6$ & 0.120(1) & 0.288(1) \\ \hline
    $L=8$ & 0.124(1) & 0.286(1) \\ \hline
    $L=10$ & 0.127(3) & 0.286(2) \\ \hline \hline
    $L \to \infty$ & 0.136(0) & $\sim 0.286$ \\ \hline
  \end{tabular}
  \caption{Results of the power-law fitting Eq.~\eqref{eq:corr_fit}. The exponent $n$ in the thermodynamic limit is estimated with weighted least squares. The prefactor $C$ is estimated to be $\sim 0.286$ in the thermodynamic limit, as the finite-size effect is negligible within the present numerical accuracy.}
  \label{tab:powerlaw}
 \end{center}
\end{table}

\subsection{\label{sec:magcata}Discussion based on magnetic catalysis}

Here, we give a brief alternative discussion for the anomalous enhancement of the Néel order in the Heisenberg model under the fictitious magnetic fields.
We will argue that it is beyond well-known magnetic catalysis for correlated systems under magnetic fields, although the spin-wave results can be understood based on this mechanism~\cite{MIRANSKY20151,PhysRevLett.73.3499,PhysRevD.52.4718,PhysRevD.46.2737,PhysRevB.85.195417,PhysRevD.107.094505,GUSYNIN1999361,PhysRevB.66.045108,JHEP07(2011)037,PhysRevResearch.2.033363, Semenoff2011}. 
Our discussion is based on the hard-core bosons which are equivalent to $S= \frac{1}{2}$ spins via the relations $c_j=S_j^-$ and $n_j=c_j^{\dagger}c_j=S_j^z + \frac{1}{2}$,
where $c_j$ is the boson annihilation operator.
The Heisenberg model at $S^z_{\rm tot}=0$ can be rewritten as the extended Hubbard model of hard-core bosons at half-filling,
\begin{align}
\mathcal{H}_{\rm B} &= t \sum_{\ev{i,j}} \left( e^{i {\theta}_{ij}} c^{\dagger}_i c_j + e^{-i {\theta}_{ij}} c^{\dagger}_j c_i \right) 
\nonumber \\
&\quad + V\sum_{\ev{i,j}}\left(n_i-\frac{1}{2}\right) \left(n_j-\frac{1}{2}\right)
\label{eq:boson}
\end{align}
with $2t=V=J$.
The hard-core condition ($n_j=0,1$) can be regarded as the strong coupling limit of the on-site Hubbard interaction $Un_j(n_j-1)$.
The boson model has particle number U(1) symmetry and anti unitary particle-hole symmetry $z\to z^{*},c_j\to c_j^{\dag}, \text{and} \, n_j\to 1-n_j$
corresponding to the $\mathrm{U}(1)\times{\mathbb Z}_2$ spin-rotation symmetry of the Heisenberg model.
The particle-hole symmetry is spontaneously broken in the charge-density wave state as in the $z$-axis Néel state of the Heisenberg model.

For the hard-core boson model, the fictitious magnetic field is nothing but a standard magnetic field for charged bosons or may be regarded as an artificial field in cold atoms~\cite{PhysRevLett.111.185301, PhysRevLett.111.185302,doi:10.1126/sciadv.1602685}.
The magnetic field leads to the Landau quantization of single-particle energies and the band structure becomes flat, which means that the hopping integral $t$ is effectively reduced.
Consequently, the interaction $Vn_in_j$ can be enhanced compared with the kinetic energy.
To be more precise, the particles are localized within a length scale $\sim l_B$, and the system size effectively becomes $\sim l_B\times l_B$ in two dimensions, where the correlation effects are amplified due to the spatial confinement.
Such a mechanism for enhanced correlation effects by magnetic fields is called magnetic catalysis and it has been extensively studied mainly for fermion systems~\cite{MIRANSKY20151,PhysRevLett.73.3499,PhysRevD.52.4718,PhysRevD.46.2737,PhysRevB.85.195417,PhysRevD.107.094505,GUSYNIN1999361,PhysRevB.66.045108,JHEP07(2011)037,PhysRevResearch.2.033363}.
Magnetic catalysis has also examined for bosons at finite temperature~\cite{ROJAS1996148,PhysRevD.86.076006}.
In this picture, it is natural to have an enhanced charge density wave order of the hard-core bosons for the Hamiltonian in Eq.\eqref{eq:boson}.

We naively expect that the magnetic length $l_B$ is a scaling variable in the present hard-core bosons as well, if the order parameter is continuous at $\phi=0$.
Indeed, in the spin-wave approximation, the Néel order parameter behaves as $\Delta m_{s}^z\sim l_B^{-\beta/\nu}=1/l_B$ , like the well-known finite-size correction $\Delta m_{s}^z\sim 1/L$~\cite{PhysRevB.39.2608,PhysRevResearch.5.013132,PhysRevB.110.054411} (see also Appendix \ref{appendix::scaling_sw}).
This is fully consistent with the physical picture of the critical magnetic catalysis seen
in mean-field approximation for $(2 + 1)$-dimensional Dirac electron systems~\cite{PhysRevResearch.2.033363}.
However, the critical behavior of the magnetic catalysis in the DMRG calculation is much stronger, $\Delta m^z_{\rm corr}\sim \phi^{0.136}=l_B^{-0.272}$, if it is continuous at $\phi=0$.
Obviously, the simple correspondence between $l_B$ and $L$ does not hold, 
and this implies that the anomalous enhancement of $\Delta m_{\rm corr}^z$ cannot be understood based on the standard magnetic catalysis.
This may be related to the fact that
the fictitious magnetic field leads not only to the Landau quantization but also to the symmetry reduction from SU(2) to $\mathrm{U}(1)\times\mathbb{Z}_2$, which is distinct from the electromagnetic field.
The bosonic nature of the constituent particles would also be important for the anomalous scaling. 

Another possibility is that $\Delta m_{\rm corr}^z$ has fluctuation-induced discontinuity at $\phi=0$, as already mentioned. 
In any case, the anomalous enhancement of the Néel order (or the charge-density wave order in terms of the bosons) is beyond the known magnetic catalysis.
It is an interesting future problem to fully clarify the origin and mechanism of the anomalous behavior of $\Delta m^z_{\rm corr}(\phi)$.

\section{\label{sec:level1} Summary}
We have studied impacts of the fictitious magnetic flux $\phi$ in the $S=\frac{1}{2}$ Heisenberg model on the square lattice.
Within the spin-wave approximation neglecting the magnon interaction, we found that the magnon excitation gap is $\Delta E\sim \sqrt{\phi}$, which can be qualitatively understood based on the dimensional analysis. 
As a result, the Néel antiferromagnetic order is enhanced by the application of a fictitious magnetic flux and it shows scaling behavior $\Delta m_{s}\sim \sqrt{\phi}$. 
This is reasonable because of the magnon gap, but in contrast with a naive expectation based on the Dzyaloshinskii-Moriya interaction appearing in the lowest-order expansion of the Aharonov-Casher phase.
We also performed the DMRG calculations to fully consider the magnon interactions and quantum fluctuations. 
Interestingly, we identified a much stronger enhancement of the Néel order with the scaling behavior $\Delta m_{\rm corr}^z\sim \phi^{0.136}$ or possibly a fluctuation-induced discontinuous enhancement.
The anomalous enhancement compared with the spin-wave approximation clearly suggests the importance of the quantum fluctuations of the Heisenberg model which is critical.
An alternative picture for the anomalous enhancement was discussed based on the magnetic catalysis for critical systems by rewriting the Heisenberg model as an interacting hard-core boson model. 
The magnetic catalysis in the present system is stronger than that in any fermion systems and is beyond the standard magnetic catalysis.
The quantum spin systems are an interesting platform for the interplay of the Landau quantization and interactions.

\begin{acknowledgements}
The DMRG calculations were performed with the use of ITensor.jl \cite{10.21468/SciPostPhysCodeb.4}. We thank Shunsuke C. Furuya and Takahiro Misawa for fruitful discussions. This study is supported by JSPS KAKENHI Grant No.22K03513.

\end{acknowledgements}

\appendix
\section{Details of the spin-wave approximation}
\label{appendix::spin_wave}
In the spin-wave approximation, we use the Holstein-Primakoff transformation~\cite{PhysRev.58.1098}. 
The annihilation (creation) operators of the hard-core bosons ${a^{(\dag)}_i, b^{(\dag)}_i}$ are introduced for the sublattices A and B. 
In a spin-$S$ system, the spin operators are written as
\begin{align}
 S^{+}_i = \sqrt{2S-a_i^{\dag}a_i} \cdot a_i, 
 \quad S^{z}_i = S - a^{\dagger}_i a_i
\end{align}
on the A sublattice and
\begin{align}
 S^{+}_i = b^{\dagger}_i\sqrt{2S-b^{\dagger}_i b_i}, 
 \quad S^{z}_i = -S + b^{\dagger}_i b_i
\end{align}
on the B sublattice. 
Then the Hamiltonian is approximated up to quadratic terms as 
\begin{align}
\mathcal{H}_{\rm sw} = JS \sum_{\ev{i,j}} \left([ a^{\dagger}_i a_i + b^{\dagger}_i b_i + e^{i \theta_{ij}} a^{\dagger}_i b^{\dagger}_j + e^{-i \theta_{ij}} a_i b_j \right],
\end{align}
where we have neglected interactions between the bosons.
The Hamiltonian in the real space basis can be straightforwardly diagonalized.
Since the spin-wave Hamiltonian has magnetic translation symmetry on a torus, physical quantities are also translationally symmetric.
It turns out that translational symmetry is kept in the bulk even for cylinder geometry.
Based on this observation,
we perform Fourier transformation in the $y$-direction in the cylinder geometry. 
The Hamiltonian is written as 
\begin{align}
\mathcal{H}_{\rm sw} &= \sum_{k_y} M^{\dagger}_{k_y} H_{k_y} M_{k_y},\\
M_{k_y} &= [ a^{\dagger}_{x_1,k_y}, a^{\dagger}_{x_2,k_y},...,b_{x_1,k_y}, b_{x_2,k_y} ]^{T}.
\end{align}
The matrix $H_{k_y}$ can be diagonalized by a Bogoliubov transformation, and physical quantities can be calculated based on the eigenvectors of $H_{k_y}$~\cite{COLPA1978327, xiao2009}.

\section{Finite-size scaling within spin-wave approximation}
\label{appendix::scaling_sw}
\begin{figure}[t]
    \centering
    \includegraphics[width=\columnwidth]{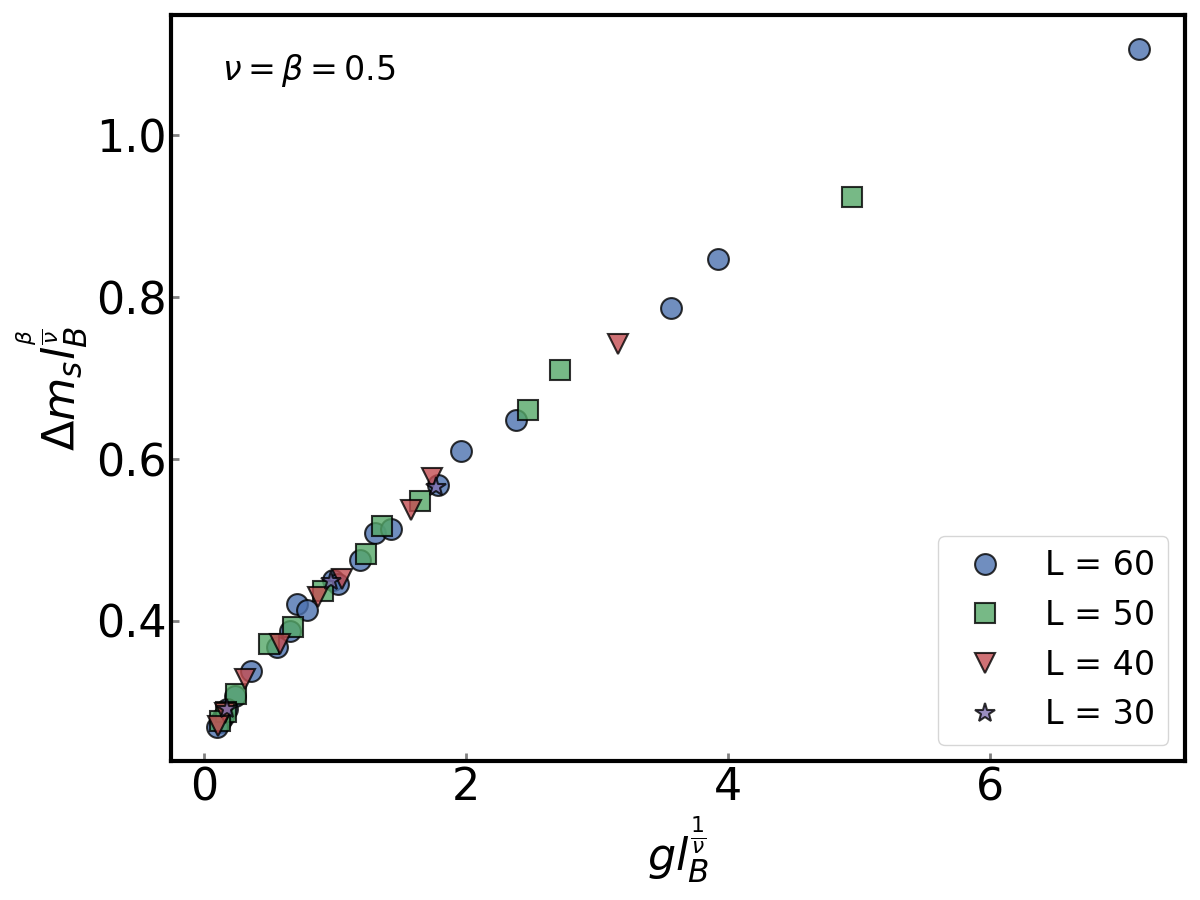}
    \caption{Scaling plot of $\Delta m_{s} (g, l_B)$ for $g = (J_z - J_{xy})/J_{xy}$ and $l_B = 1/ \sqrt{\phi}$.
    The parameters are $0 \leqq \phi \lesssim 0.01$ and $0 \leqq g \lesssim 0.01$. The critical exponents are fixed as $\beta=\nu=1/2$. 
    }
    \label{fig:scaling}
\end{figure}
\begin{figure}[t]
    \centering
    \includegraphics[width=\columnwidth]{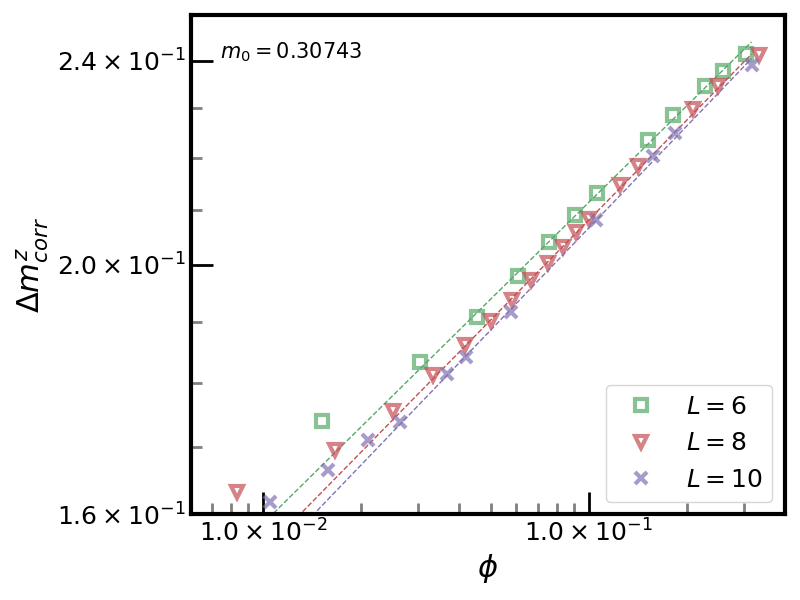}
    \caption{Scaling plot of $\Delta m_{\rm corr}^z (\phi)$. The zero field value is $m_{\rm corr}^0=0.30743$ obtained in the previous quantum Monte Carlo study~\cite{PhysRevB.82.024407}. The dashed lines are power law fitting. }
    \label{fig:log-log_QMC_previous}
\end{figure}
To establish that the magnetic length is the dominant cut off scale in the Heisenberg model under the fictitious magnetic field within the spin-wave approximation, we extend our discussion to the $XXZ$ model and perform a finite-size analysis.
We consider the $XXZ$ model under a fictitious magnetic field,
\begin{align}
\mathcal{H} = \sum_{\ev{i,j}} \left[ \frac{J_{xy}}{2} \left( e^{i {\theta}_{ij}} S^{+}_i S^{-}_j + e^{-i {\theta}_{ij}} S^{-}_i S^{+}_j \right) + J_z S^z_i S^z_j \right],
\end{align}
where $J_z / J_{xy} \geqq 1$. 
The Heisenberg point $J_{z} = J_{xy}$ is SU(2) symmetric and is located at a bicritical point~\cite{PhysRevB.16.2191}, whereas the ground state is Néel ordered along the $z$-axis under the condition  $J_{z} > J_{xy}$.
Near the critical point $J_{z} = J_{xy}$, we suppose that the difference $\Delta m_s$ is described by a scaling function $f$, 
\begin{align}
    \Delta m_s (g, l_B) = l_B^{-\beta/\nu} f(g l_B^{1/\nu} )
\end{align}
where $g = \left(J_z - J_{xy} \right)/J_{xy}$ and $\beta$, $\nu$ are the critical exponents at $\phi=0$ for the difference $\Delta m_s(g, l_B \to \infty ) \sim g^{\beta}$ and the correlation length $\xi(g, l_B \to \infty ) = g^{-\nu}$.  
Note that we use $l_B$ instead of $L$ for the finite-size scaling analysis since $l_B$ is considered the dominant length scale rather than the system size $L$ when $l_B<L$.
In previous works at $\phi=0$, the spin-wave approximation gives the enhancement of the Néel order $\Delta m_s(J_{z}) \propto \sqrt{J_z - J_{xy}}$ and energy gap $\Delta E \propto \sqrt{J_z-J_{xy}}$ in the leading order~\cite{PhysRevB.44.11869,PhysRevB.46.6276}. 
The same critical behaviors have also been obtained by the extensive numerical calculations~\cite{PhysRevResearch.5.013132,PhysRevB.110.054411}.
These studies give the critical exponents $\beta = 0.5$ and $\nu = 0.5$ under the assumption that the dynamical critical exponent is $z=1$ since the correlation length $\xi$ scales as $\xi^z \sim \frac{1}{\Delta E}$.
Figure~\ref{fig:scaling} shows the finite-$l_B$ scaling with the fixed critical exponents $\beta=\nu= \frac{1}{2}$. 
We see that the numerical data successfully collapse onto a single curve,
and the finite-$l_B$ scaling indeed holds.
Thus the two length scales $l_B$ and $L$ are interchangeable within the spin-wave approximation, which may be reasonable.
However, this equivalence is broken in the DMRG results where the full quantum fluctuations are considered as discussed in the main text.

\section{Robustness of power-law fitting for $m^0_{\rm corr}$}
\label{appendix::QMC and DMRG}

In the main text, the zero-field value of the Néel magnetization $m_{\rm corr}^0=0.2985$ is used for the discussions of $\Delta m_{\rm corr}^z$.
To check the robustness of our discussion, we perform the same power-law fitting using a different zero-field value $m^0_{\rm corr} = 0.30743$ obtained in the previous quantum Monte Carlo study~\cite{PhysRevB.82.024407}.
Figure~\ref{fig:log-log_QMC_previous} shows the logarithmic scale plot of $\Delta m^z_{\rm corr}$ vs $\phi$ using $m_{\rm corr}^0=0.30743$, similar to Fig.~\ref{log-log}. 
The power-law behavior can be seen for $\phi>\phi_0$, where the crossover scale is $\phi_0 = 0.05$. 
The calculation results are summarized in Table \ref{table:log-log_QMC_previous}. 
The exponent $n$ and the prefactor $C$ are extrapolated to $n \simeq 0.141, C\simeq 0.275$, and they are close to those obtained in the main text. 
These results indicate that the slight deviation of the background $m_{\rm corr}^0$ is negligible in the power-law fitting.
\begin{table}[thb]
 \centering
 \begin{center}
  \begin{tabular}{|c|c|c|}
     \hline
    System size $L$ & Exponent $n$& Prefactor $C$\\ \hline
    $L=6$ & 0.124(2) & 0.281(1) \\ \hline
    $L=8$ & 0.128(1) & 0.280(1) \\ \hline
    $L=10$ & 0.131(3) & 0.279(2) \\ \hline \hline
    $L \to \infty$ & 0.141(1) & 0.275(3) \\ \hline
  \end{tabular}
  \caption{Results of power-law fitting Eq.~\eqref{eq:corr_fit} with $m_{\rm corr}^0=0.30743$. The values of $n$ and $C$ in the thermodynamic limit are evaluated by linear-fitting with weighted least squares.}
  \label{table:log-log_QMC_previous}
 \end{center}
\end{table}
\nocite{*}

\bibliography{apssamp}

\end{document}